\documentclass[aps,pre,groupedaddress,showpacs,preprint]{revtex4} 
\usepackage[dvips]{epsfig}
\usepackage{dcolumn}
\usepackage{subfigure}%
\usepackage{bm}
\usepackage{amssymb}
\usepackage{amsmath}
\usepackage{color,soul}

\usepackage[latin1]{inputenc}   

\usepackage{color,soul}  
\setstcolor{red}         

\newcommand{\be}{\begin{equation}}
\newcommand{\ee}{\end{equation}}
\newcommand{\ba}{\begin{eqnarray}}
\newcommand{\ea}{\end{eqnarray}}
\newcommand{\besu}{\begin{subequations}}
\newcommand{\esu}{\end{subequations}}

\newcommand{\yb}{\mathbf{y}}
\newcommand{\xb}{\mathbf{x}}
\newcommand{\zb}{\mathbf{z}}
\newcommand{\ub}{\mathbf{u}}
\newcommand{\vb}{\mathbf{v}}
\newcommand{\fb}{\mathbf{f}}
\newcommand{\Fb}{\mathbf{F}}
\newcommand{\epsilonb}{\boldsymbol{\epsilon}}

\begin{document}
\title{Diffusion in an expanding medium: Fokker-Planck equation, Green's function and first-passage properties}
\author{S. B. Yuste$^{1}$, E. Abad$^{2}$, C. Escudero$^{3}$}
\affiliation{
 $^{1}$ Departamento de F\'{\i}sica and Instituto de Computaci\'on Cient\'{\i}fica Avanzada (ICCAEX) \\
 Universidad de Extremadura, E-06071 Badajoz, Spain \\
 $^{2}$ Departamento de F\'{\i}sica Aplicada and Instituto de Computaci\'on Cient\'{\i}fica Avanzada (ICCAEX) \\ Centro Universitario de M\'erida \\ Universidad de Extremadura, E-06800 M\'erida, Spain \\
$^{3}$
Departamento de Matem\'aticas \\
Universidad Aut\'onoma de Madrid \\
and Instituto de Ciencias Matem\'aticas \\
Consejo Superior de Investigaciones Cient\'{\i}ficas,\\
E-28049 Madrid, Spain
}

\begin{abstract}

We present a classical, mesoscopic derivation of the Fokker-Planck equation for diffusion in an expanding medium. To this end, we take a conveniently generalized Chapman-Kolmogorov equation as the starting point. We obtain an analytical expression for the Green's function (propagator) and investigate both analytically and numerically how this function and the associated moments behave.  We also study first-passage properties in expanding hyperspherical geometries. We show that in all cases the behavior is determined to a great extent by the so-called Brownian conformal  time $\tau(t)$, which we define via the relation $\dot \tau=1/a^2$, where $a(t)$ is the expansion scale factor.  If the medium expansion is driven by a power law [$a(t) \propto t^\gamma$ with $\gamma>0$], we find interesting crossover effects in the mixing effectiveness of the diffusion process when the characteristic exponent $\gamma$ is varied. Crossover effects are also found at the level of the survival probability and of the moments of the first passage-time distribution with two different regimes separated by the critical value $\gamma=1/2$. The case of an exponential scale factor is analyzed separately both for expanding and contracting media. In the latter situation, a stationary probability distribution arises in the long time limit.

\end{abstract}


\maketitle

\section{Introduction}

The overwhelming majority of studies devoted to diffusion processes assume them to take place in static media. However, expanding (or contracting) media are by no means a rarity in Nature. In fact, we live in an expanding Universe~\cite{Riess1998,Perlmutter1999}, and elementary biological processes such as morphogenesis (i.e., the process whereby a living being evolves from a single cell to a fully developed adult) involve tissue expansion processes. Moreover, in a number of cases,  physical processes, and diffusion processes in particular, are significantly affected by the expansion or contraction of the media in which they take place. For instance, in developmental biology it is well-known that the formation of biological structures via diffusion-mediated processes can be significantly altered by the concomitant growth of tissues and organs~\cite{Crampin1999,Crampin2001,Crampin2002,Simpson2015a}. Another example, taken from Cosmology, is the diffusion of cosmic rays in the expanding Universe~\cite{Berezinsky2006,Berezinsky2007,Aloisio2009}; moreover, the general problem of a fluid diffusing in the expanding Universe was addressed in~\cite{Haba2014}, and this in fact could be considered a simplified model for the evolution of the Universe itself. All these facts highlight the necessity of developing a stochastic theory able to address the dynamics of ensembles of random walkers embedded in an expanding space by conveniently bridging the gap between the mesoscopic and the macroscopic level of description. The present paper is a step in this direction.

To the best of our knowledge,  the derivation of the classical diffusion equation for transport in growing media has been carried out via two possible pathways.  The first one uses mass conservation arguments together with the assumption that the particle flux is proportional to the concentration gradient (Fick's first law) to obtain a generalized diffusion equation (Fick's second law) \cite{Crampin1999,Berezinsky2006}. The second approach relies on a coarse-grained stochastic model, implying that the medium is first partitioned into boxes, and then a master equation formalism describing fluxes between neighboring boxes is employed to obtain the generalized diffusion equation \cite{Baker2010,Yates2012}.
In the present paper, we shall follow ``Einstein's footsteps'' and develop an alternative description based on a random walk model. In our case, the $d$-dimensional Fokker-Planck equation describing transport in a growing medium is obtained from the corresponding Chapman-Kolmogorov equation. This is done in Sec.~\ref{secmderivation}. Following this, in Sec.~\ref{secGruex} we compute the Green's function (propagator) $P(\yb,t)$ for the case of a uniform expansion. The propagator is expressed in terms of the Brownian conformal time $\tau=\tau(t)$, defined by means of the differential equation $\dot \tau=1/a^2(t)$, where $a(t)>0$ stands for the expansion scale factor. The specific time dependence of the latter turns out to have a strong influence on the manner in which particles spread, which can be better characterized with the help of some definitions introduced in Sec.~\ref{secDiffPulse}. The underlying phenomenology is discussed in Sec.~\ref{secPowerlawexp}  on the basis of a specific, yet important  example, namely, the case of a power-law scale factor $a(t)\propto t^\gamma$. In this context, a rich behavior is seen to emerge as the characteristic exponent $\gamma$ is varied; see~\cite{ce1,ce2,ce3,ce4,ce5,ce6} for a variety of physical systems displaying a similar behavior.
Comparison with stochastic simulations is provided, and the corresponding moments $\langle y^n \rangle$ are also evaluated
(for the one-dimensional case as well as for the higher dimensional case).
In this context, we find that diffusion in an expanding space is non-stationary and non-ergodic in a way similar to scaled Brownian motion, a Gaussian approximation for Continuous Time Random Walks which is widely used for fitting experimental particle trajectories displaying anomalous diffusion.
In Sec.~\ref{expexp} we analyze the case of an exponential scale factor, both for expanding and for contracting media. In the latter case we find that the system converges to a stationary probability distribution, a phenomenon that does not take place in contracting media driven by power law scale factors.

Finally, in Sec.~\ref{secSurvival},  we consider diffusion problems in expanding media with absorbing boundaries. Such problems are often taken as the starting point to compute a number of characteristic first-passage properties, e.g. survival probabilities and moments of the first-passage time distribution. In turn, these quantities play a central role in the classical theory of diffusion-controlled reactions, and more specifically in so-called target and trapping problems. We again find interesting crossover effects when the medium expansion is driven by a power law and its characteristic exponent is varied. Before moving on to the derivation of our main results, we take the opportunity to highlight the fundamental difference between the first-passage problem addressed here and a widely studied class of problems concerning systems with moving boundaries~\cite{Tuckwell1984} (e.g. absorption of a diffusing particle at the boundaries of an expanding cage, a receding wall, etc.). In the  latter case, physical distances are stationary, and only the position of the system boundaries changes in time.
Our main conclusions are stated in Sec.~\ref{conclusions}, where we also outline a series of open questions.

\section{Mesoscopic derivation of the diffusion equation}
\label{secmderivation}

\subsection{Implementing volume expansion}
Let $\mathbf{y}(t)$ denote the position at time $t$ of a point particle with no motion of its own. If the embedding medium shrinks or expands (in what follows, and without loss of generality, we say ``expands'' for brevity), the particle will experience a drift, as a result of which its position at a later time $t'$ will be different,  $\mathbf{y}'=\mathbf{y}(t')=\mathbf{F}(\mathbf{y},t,t'-t)$. A suitable way to describe the medium expansion consists in expressing the Lagrangian coordinate (or physical distance) $\mathbf{y}$ in terms of the Eulerian coordinate (or comoving distance \cite{Berezinsky2006}) $\mathbf{x}$:
\begin{equation}
\mathbf{y}(t)=\mathbf{f}(\mathbf{x},t).
\end{equation}
At any time $t$,  $\mathbf{f}$ is a continuous bijective function of $\mathbf{x}$ with the property $\mathbf{x}=\mathbf{f}(\mathbf{x},t_0)$,  where $t_0$ is the initial time. The latter is taken to be the instant when the observation of the particle's motion begins.   In particular
\begin{align}
\mathbf{y}(t+\Delta t)&=\mathbf{F}(\mathbf{y},t) = \mathbf{f}(\mathbf{x},t+\Delta t)   \\
 &=  \mathbf{y}(t)+\ub(\mathbf{x},t)  \Delta t +O(\Delta t)^2,
 \label{yTay}
\end{align}
where $\mathbf{u}(\xb,t)\equiv\dot{\mathbf{f}}(\xb,t)\equiv  \partial \mathbf{f}/\partial t$ is the expansion velocity field, and the short-hand notation $\mathbf{F}(\mathbf{y},t)\equiv\mathbf{F}(\mathbf{y},t,\Delta t)$ has been used.  Later on we shall denote the function $\mathbf{u}[\xb(\yb,t),t]=\mathbf{u}[\fb^{-1}(\yb,t),t]$ by $\mathbf{u}(\yb,t)$, and so the function under consideration will be distinguished solely by the symbol $\xb$ or $\yb$  used in the argument.

Due to the medium expansion a $d$-dimensional volume $\Delta V$ centered at $\yb$ at time $t$ evolves into a volume $\Delta'V$ centered at $\yb'=\mathbf{F}(\mathbf{y},t)$ at time $t+\Delta t$. The ratio between these two volumes is simply the determinant of the Jacobian $J$ associated with the expanding  transformation
\begin{equation}
\label{ratioVol}
\frac{\Delta' V}{\Delta V}=|J(\mathbf{y},t,\Delta t)|=
\left| \frac{\partial (y'_1,y'_2,\cdots,y'_d)}
{\partial (y_1,y_2,\cdots,y_d)} \right|\equiv
\left| \frac{\partial (F_1,F_2,\cdots,F_d)}
{\partial (y_1,y_2,\cdots,y_d)} \right|.
\end{equation}
From Eq.~\eqref{yTay} one sees that
\begin{equation}
J(\mathbf{y},t,\Delta t)=\mathbb{I} + \frac{\partial(u_1,u_2,\cdots,u_d)}{\partial(y_1,y_2,\cdots,y_d)} \Delta t+O(\Delta t)^2,
\end{equation}
and therefore
\begin{equation}
\label{ratioV}
\frac{\Delta' V}{\Delta V}= |J(\mathbf{y},t,\Delta t)|=
1+  \nabla\cdot\ub\, \Delta t+O(\Delta t)^2.
\end{equation}
This expression shows that
\begin{equation}
\label{nafb}
\nabla\cdot\ub=\lim_{\Delta t \to 0} \frac{\Delta'V/\Delta V-1}{\Delta t}=\lim_{\Delta t \to 0} \frac{\Delta V(t+\Delta t)/\Delta V(t)-1}{\Delta t}
\end{equation}
is simply the relative volume expansion rate. For the particular case of an expansion which is anisotropic but homogeneous in each Cartesian direction one has
\begin{equation}
\label{yax}
y_i=f_i(\xb,t)=a_i(t)x_i,
\end{equation}
with $u_i(\xb,t)=\dot f_i(\xb,t)=\dot a_i(t) x_i$ and $f_i^{-1}(\yb,t)=x_i=y_i/a_i(t) $. Then,
\begin{equation}
\label{dotf}
u_i(\yb,t)= \dot f_i[\fb^{-1}(\yb,t),t]=\frac{\dot a_i}{a_i}\,  y_i
\end{equation}
and $\nabla\cdot\ub=\sum_{i=1}^d \dot a_i/a_i$.   For the case of uniform expansion $a_i(t)=a(t)$ we can write $\nabla\cdot\ub=H\,d$, where $H(t)=\dot a/a$ is the Hubble parameter and $a(t)$ is the scale factor \cite{Berezinsky2006,Knobloch2015}.

\subsection{Generalized Chapman-Kolmogorov equation}
Our next step consists in superimposing an intrinsic stochastic particle motion to the extrinsic (deterministic) motion caused by the medium expansion (the words ``walker'' and ``particle'' will be used as synonyms in what follows). To this end, we hereafter adopt a mesoscopic description of the diffusion process in terms of a random walk approach. In this framework, we consider a collection of particles taking steps of variable size $z$ at discrete times $t_m$ separated by constant intervals $\Delta t=t_{m+1}-t_m$, whereby we keep in mind the idea of eventually letting $\Delta t$ shrink to zero (note, however, that the steps are considered to be instantaneous, i.e., they occur on a much shorter time scale than the waiting time between consecutive steps).

Let $P(\yb,t^+_n|\yb_0,t_0)$ [$P(\yb,t^-_n|\yb_0,t_0)$] be the probability density to find a walker in an infinitesimal volume about $\yb$ at time $t^+_n$ [$t^-_n$], (i.e., immediately \emph{after} [\emph{before}] taking the $n$-th step at time $t_n=t_0+n\Delta t$), given the walker's initial position $\yb_0\equiv \yb(t_0)$. One can now take advantage of the Markovian character of the walker's motion to obtain the following version of the Chapman-Kolmogorov equation:
\begin{align}
\label{CK31}
P(\yb,t_{n+1}^+|\yb_0,t_0)&=\int P(\yb-\zb,t_{n+1}^-|\yb_0,t_0) P(\yb,t_{n+1}^+|\yb-\zb,t_{n+1}^-) \, d\zb.
\end{align}
Using a simplified notation, this equation can be written as follows:
\begin{align}
\label{ecuCKa}
P^+_{n+1}(\yb)&=\int P^-_{n+1}(\yb-\zb) \,p(\zb|\yb-\zb,t_{n+1}) \, d\zb,
\end{align}
where $P^+_{n}(\yb)\equiv P(\yb,t^+_n|\yb_0,t_0)$,  $P^-_{n}(\yb)\equiv P(\yb,t^-_n|\yb_0,t_0)$, and $p(\zb|\yb,t_{n})\equiv P(\yb+\zb,t_{n}^+|\yb,t_{n}^-)$ denotes the probability that the  $n$-th step of  the walker situated at position $\yb$ at time $t_n^-$  results in a displacement $\zb$.  A walker located at $\yb$ immediately \emph{after} its $n+1$-th step (taken at time $t_{n+1}$) may have reached its location from any other position $\yb-\zb$ (occupied by the walker immediately \emph{before} taking the $n+1$-th step) by means of an appropriate single-step displacement $\zb$. Equations~\eqref{CK31}-\eqref{ecuCKa} simply state that the probability of the walker being  at $\yb$ at time  $t_{n+1}^+$ is  equal to the sum of the infinitesimal contributions stemming from all its possible previous positions.

\begin{figure}[t]
\begin{center}
 \includegraphics[width=0.5\textwidth,angle=0]{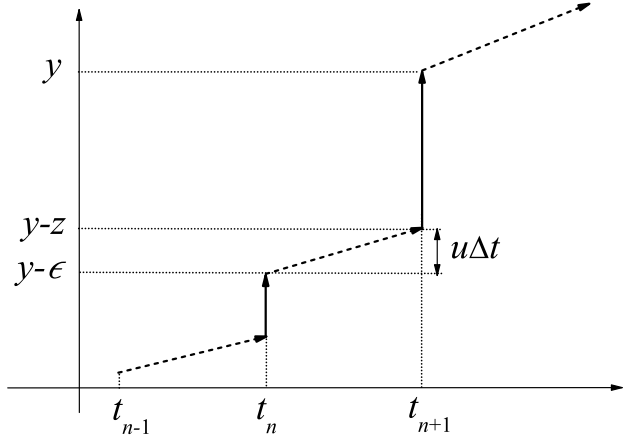}
\end{center}
\caption{\label{FigEsqExp} Schematic picture of the random walker's motion due to the combined action of random steps (solid arrows)  and the deterministic drift arising from the medium expansion (dotted arrows). Recall that $\yb'=\yb-\zb=\Fb(\yb-\epsilonb,t_n)$.}
\end{figure}

In the case where the walker does not move during the time interval $t_n<t<t_{n+1}$,  one has $P_n^+(\yb)=P_{n+1}^-(\yb)$; this then means that one has a static (non-growing) medium, and the Chapman-Kolmogorov equation \eqref{ecuCK1} immediately reduces to the standard one:
\begin{align}
\label{ecuCK1}
P^+_{n+1}(\yb)&=\int P^+_{n}(\yb-\zb)\, p(\zb|\yb-\zb,t_{n+1}) \,d\zb.
\end{align}
However, for expanding media, the probability density to find a walker lacking intrinsic motion inside a volume $\Delta V$  about position $\yb$ at time $t$ is different from the probability density to find the walker inside a volume $\Delta ' V$ about its new position $\yb'=\Fb(\yb,t)$ at time $t+\Delta t$ because of the change in volume brought about by the medium expansion (the latter introduces a dilution effect at the level of the particle concentration).  However, for such a walker the \emph{probability} of sojourn in a given volume is conserved in the course of the expansion, i.e.,
\begin{equation}
\label{PVpV}
 P(\yb',t+\Delta t)  \Delta' V = P(\yb,t) \Delta V.
\end{equation}
In particular, if we take $\yb'=\yb-\zb$ and $t=t_n$ the above conservation relationship leads to the following equation:
\begin{align}
\label{Pmm}
 P_{n+1}^-(\yb-\zb)= P_{n}^+(\yb-\epsilonb) \frac{\Delta V}{\Delta' V} =
\frac{P_{n}^+(\yb-\epsilonb))}{|J(\yb-\epsilonb,t_n,\Delta t)|},
\end{align}
where  Eq.~\eqref{ratioV}  has been used. The parameter $\epsilonb$ is defined via the relation
\begin{equation}
\label{epsDef}
\yb-\zb=\Fb(\yb-\epsilonb,t_{n})
\end{equation}
(see Fig.~\ref{FigEsqExp}).
Inserting Eq.~\eqref{Pmm} into Eq.~\eqref{ecuCKa} one obtains the final form of the Chapman-Kolmogorov equation for growing media:
\begin{equation}
\label{CK14}
P_{n+1}^+(\yb)=\int \frac{P_{n}^+(\yb-\epsilonb)}{|J(\yb-\epsilonb,t_n,\Delta t)|}\; p(\zb|\yb-\zb,t_{n+1}) d\zb.
\end{equation}
Note that this equation differs from the standard one for non-growing media, Eq.~\eqref{ecuCK1}, by (i) the Jacobian term, which is due to the medium expansion [cf.~Eq.~\eqref{ratioVol}], and (ii) the change of $\tau$ by $\epsilon$ in the argument of $P_n^+$, representing the change in the physical coordinate $\yb$ of a point that arises merely from the drift induced by the medium expansion (see discussion in Sec.~\ref{secFPex}).

\subsection{Fokker-Planck equation describing random motion in an expanding medium}
\label{secFPex}
In order to obtain the relevant Fokker-Planck (FP) equation we now expand the integrand of Eq.~\eqref{CK14} in a Taylor series about the point $(\yb,t_n)$.  In doing so, we take into account that both $\zb$ and $\epsilonb$ are small, the difference $\zb-\epsilonb$ being of the order of $\Delta t$. The latter statement can be easily proven by using Eq.~\eqref{yTay} in the definition of $\epsilonb$ [cf. Eq.~\eqref{epsDef}]. One is then left with the following equation:
\begin{equation}
\label{epsTay}
\epsilonb=\zb+\ub(\xb,t_n) \Delta t+o(\Delta t),
\end{equation}
where the rest $o(\Delta t)$ is a sum of terms which all go to zero faster than $\Delta t$.

Let us denote by $\text{Tay}[G,\boldsymbol{\xi};\yb,t]$ the Taylor expansion of the function $G$ in powers of  $\xi$ about the point $(\yb,t)$. Then we have
\begin{align}
\label{TayFloat}
\text{Tay}\left[  \frac{P_{n}^+(\yb-\epsilonb)}{|J(\yb-\epsilonb,t_n,\Delta t)|},\epsilonb;\yb,t \right]=&
\text{Tay}\left[  \frac{P_{n}^+(\yb-\zb)}{|J(\yb-\zb,t_n,\Delta t)|},\zb;\yb,t \right]  \nonumber\\
&-\Delta t \sum_{i=1}^d u_i(\xb,t_n) \frac{\partial}{\partial y_i}  \frac{P_{n}^+(\yb)}{|J(\yb,t_n,\Delta t)|} +o(\Delta t).
\end{align}
Note that the Taylor expansion on the left hand side differs from that on the right hand side because $\epsilonb$ and $\zb$ are not the same. Physically, this is due to the fact that the displacement of the particle from $\yb-\epsilonb$ to $\yb-\zb$ is solely induced by the drift associated with the medium expansion (cf. Fig. ~\ref{FigEsqExp}). In terms of the velocity field, this displacement is simply expressed as $\ub(\xb,t_n) \Delta t+o(\Delta t)$. On the other hand, one has
\begin{align}
\text{Tay}\left[ p(\zb|\yb-\zb,t_{n+1}),\zb;\yb,t_n \right]=&
\text{Tay}\left[ p(\zb|\yb-\zb,t_{n}),\zb;\yb,t_n \right]  +\Delta t \frac{\partial}{\partial t} p(\zb|\yb,t_{n}) +o(\Delta t).
\end{align}
Hence, denoting by $I$ the integrand on the right hand side of Eq.~\eqref{CK14}, one obtains
\begin{align}
\label{Tay12}
I=&\text{Tay}\left[  \frac{P_{n}^+(\yb-\zb)\, p(\zb|\yb-\zb,t_{n})}{|J(\yb-\zb,t_n,\Delta t)|},\zb;\yb,t \right]
-\Delta t \sum_{i=1}^d u_i(\xb,t_n) \frac{\partial}{\partial y_i}  \frac{P_{n}^+(\yb)}{|J(\yb,t_n,\Delta t)|} \nonumber \\
& +\Delta t \frac{\partial}{\partial t} p(\zb|\yb,t_{n}) +o(\Delta t).
\end{align}
Inserting this expression into Eq.~\eqref{CK14} and taking into account that   $\int  \partial p(\zb|\yb,t_{n})/{\partial t} \, d\zb=0$, one finds
\begin{align}
\label{FKo1}
P_{n+1}^+(\yb) &= \sum_{n=0}^\infty T_n  -\Delta t \sum_{i=1}^d u_i(\xb,t_n) \, \int p(\zb|\yb,t_{n})\;\frac{\partial}{\partial y_i}  \frac{P_{n}^+(\yb)}{|J(\yb,t_n,\Delta t)|} \,d\zb+o(\Delta t),
\end{align}
where $T_n$ stands for the integral over $\zb$ of the product of $p(\zb|\yb,t_{n})$ and the  $n$-th order term of the Taylor expansion in Eq.~\eqref{Tay12}.  In particular, one has
\begin{align}
T_0=&\int \frac{P_{n}^+(\yb)}{|J(\yb,t_n,\Delta t)|}  p(\zb|\yb,t_{n}) d\zb
=  P_{n}^+(\yb)  +\left[\frac{1}{|J(\yb,t_n,\Delta t)|} -1 \right] P_{n}^+(\yb),
  \label{T0}\\
T_1=&-\sum_{i=1}^d \frac{\partial}{\partial y_i} \left\{ \frac{P_{n}^+(\yb)}{|J(\yb,t_n,\Delta t)|} \int z_i
 \; p(z|\yb,t_{n}) d\zb\right\},
 \label{T1}\\
T_{2}=&\sum_{i,j=1}^d \frac{\partial^2}{\partial y_i \partial y_j}  \left\{ \frac{P_{n}^+(\yb)}{|J(\yb,t_n,\Delta t)|}\; \int z_i z_j
  p(\zb|\yb,t_{n}) d\zb\right\}.
  \label{T2}
\end{align}
Note that the second term on the right hand side of  Eq.~\eqref{T0} appears only because $ |J(\yb,t_n,\Delta t)|\neq 1$. This term represents the decrease of the particle density (\emph{dilution}) due to the medium expansion [see Eqs.~\eqref{PVpV} and \eqref{Pmm}, as well as the discussion preceding those two equations].  Using Eq.~\eqref{ratioV} in Eq.~\eqref{T0}, one finds
\begin{align}
T_0&=  P_{n}^+(\yb)  -P_{n}^+(\yb)\,\nabla\cdot\ub \,\Delta t  +o(\Delta t).
  \label{T0Exp}
\end{align}

In order to take the limit $\Delta t\to 0$ we shall use the more convenient notation $P(\yb,t_n)\equiv P_{n}^+(\yb)$, leading to the corresponding time derivative $\lim_{\Delta t \to 0} [P_{n+1}^+(\yb)-P_{n}^+(\yb)]/\Delta t=\partial P(\yb,t)/\partial t$.  Inserting Eqs.~\eqref{T1}-\eqref{T0Exp} into Eq.~\eqref{FKo1} and taking the limit $\Delta t\to 0$ in the resulting expression, one obtains
\begin{align}
\frac{\partial }{\partial t} P(\yb,t)  &=
-\; P(\yb,t) \sum_{i=1}^d \frac{\partial u_i}{\partial y_i}
- \sum_{i=1}^d  u_i\,  \frac{\partial }{\partial y_i} P(\yb,t)
\nonumber \\
 &
-\sum_i  \frac{\partial }{\partial y_i}  A_i(\yb,t) P(\yb,t)
 +  \sum_{i j} \frac{\partial^2  }{\partial y_i \partial y_j}   D_{ij}(\yb,t)  P(\yb,t)
+\lim_{\Delta t\to 0} \sum_{n=3}^\infty \frac{T_n}{\Delta t}  \label{FK1},
\end{align}
where
\begin{align}
\label{Ai}
A_i(\yb,t)&\equiv v_i(\yb,t)= \lim_{\Delta t\to 0} \frac{\int z_i \; p(\zb|\yb,t) d\zb}{\Delta t},\\
D_{ij}(\yb,t)&= \lim_{\Delta t\to 0} \frac{ \int z_i z_j  p(\zb|\yb,t_{n}) d\zb }{2 \Delta t}.
\label{Dij}
\end{align}
Note that the first two terms on the right hand side of Eq.~\eqref{FK1} can be written as $\nabla\cdot (\ub P)$. Finally, the limits of Eqs.~\eqref{Ai}-\eqref{Dij} exist and  $\lim_{\Delta t\to 0}  T_n/\Delta t=0$ for $n\ge 3$ if we assume that   $p(\zb|\yb,t)$ has the characteristic properties of a continuous Markov process (see, for example, \cite{Guillespie1996} or  Secs.~3.4 in \cite{Gardiner} and 7.4 in \cite{Gillespie2013}). Under the above assumption,  Eq.~\eqref{FK1} (a kind of forward Kramers-Moyal expansion) yields the following FP equation:
\begin{align}
\frac{\partial }{\partial t} P(\yb,t)  &=
-\sum_i  \frac{\partial }{\partial y_i}
\left[u_i(\yb,t)+A_i(\yb,t)\right] P(\yb,t)
 +  \sum_{i j} \frac{\partial^2  }{\partial y_i \partial y_j}   D_{ij}(\yb,t)  P(\yb,t).
  \label{FKfinal}
\end{align}
Eq.~\eqref{FKfinal} is the cornerstone of our analysis in subsequent sections.
The interpretation and properties of $A_i\equiv v_i$ (drift vector) and $D_{ij}$ (diffusion coefficient matrix) are similar to those of the analogous quantities appearing in the standard FP equation describing transport in a static medium \cite{Gardiner}. In the above equation two different drift velocities appear, namely, an ``intrinsic'' drift velocity $\vb$ arising from the asymmetry of the jump length PDF $p(\zb|\yb,t)$ of the random walker, and an ``extrinsic'' drift velocity $\ub$ exclusively due to the expansion of the embedding medium.
Of course, if there is no expansion, $a_i(t)=1$, $\ub=0$, and one recovers the FP equation for a static medium.

Note that Eq.~\eqref{FKfinal} is simply the standard FP equation augmented with a term $-\nabla\cdot (\ub P)$ describing the effect of the medium expansion. This additional term can be split into the drift term  $-(\ub \cdot \nabla) P$ and the dilution term $-P (\nabla\cdot \ub)$ [see the discussions following Eqs.~\eqref{TayFloat} and \eqref{T2}]. For the special case of the expansion given by  Eq.~\eqref{yax}, $y_i=f_i(\xb,t)=a_i(t)x_i$,  this extra term simply becomes [see the discussion after Eq.~\eqref{nafb}]
\begin{equation}
\nabla\cdot (\ub P)=  \sum_{i=1}^d  \frac{\dot a_i}{a_i} \frac{\partial }{\partial y_i} [y_i  P(\yb,t)].
\end{equation}

\subsection{Langevin equation for transport in an expanding medium}

Using standard procedures  \cite{Gardiner,Gillespie2013}, one can show that Eq.~\eqref{FKfinal} is equivalent to the following set of  Langevin equations:
\begin{align}
dy_i  &= \left[u_i(\yb)+A_i(\yb,t)\right] dt + \sqrt{2} \, \sum_{j=1}^d b_{ij}(\yb,t) dW_j(t),
\end{align}
where $\{W_j(t)\}_{j=1}^d$ is a family of independent Wiener processes, $dW_i(t)$ stands for the increment of the $i-$th Wiener process at time $t$, and $D_{ij}\equiv \sum_{k=1}^d b_{ik}b_{kj}$.
From Eq.~\eqref{FKfinal} it is obvious that the It\^{o} interpretation of stochastic calculus is being used.
For the one-dimensional case there is a single coefficient $b$, which is univocally determined by the diffusion coefficient $D(y,t)$. In this particular case, the above Langevin equation takes the form
 \begin{align}
y(t+dt)=y(t) + \left[u+A(y,t)\right] dt + \sqrt{2 D(y,t)} \, dW(t).
\end{align}
The interpretation of this equation is straightforward: the walker moves ballistically with a drift velocity $u+A(y,t)$, on which Gaussian fluctuations are superimposed.  A version of this equation in terms of finite differences will be used in Sec.~\ref{secPowerlawexp} to simulate free diffusion in a medium whose expansion is described by Eq.~\eqref{yax}.

We have restricted ourselves to the overdamped case, but a natural extension of our work would consist in incorporating a mass term into the equation of motion. A full solution of this problem is beyond the scope of the present work, but a short qualitative discussion is still possible. For the underdamped case, in the absence of the medium expansion, it is well known that a crossover takes place from a ballistic regime valid for short times to a diffusive regime at longer times. Both regimes are separated by a typical crossover time $t_\times$. For static media,  $t_\times$ is related to the so called characteristic diffusion length $\ell_d$ by the equation $\ell _d^2=D(t_\times-t_0)$ \cite{Guillespie1996}. For a growing medium, this distance, $\ell_d^e$, expands by a factor $a(t)$, so that $\ell _d^e=a(t) \ell_d$, leading to  $t_\times^e-t_0= a^2(t) (t_\times-t_0)$  with $ t_0\le t\le t_\times^e$. For standard microscopic systems $\ell_d$ and $t_\times-t_0$ are very small, and hence $\ell _d^e $ and $t_\times^e-t_0$ is also small if one assumes that $a(t)$ does not change very significantly in this time interval. In other words, for standard expanding media, one expects a negligible effect on the typical crossover time separating the ballistic regime from the diffusive regime. Of course, this simple  argument needs confirmation by more rigorous analysis.

\section{Propagator and moments for the case of uniform expansion}
\label{sec:ict}

\subsection{Generic results}
\label{secGruex}

The solution of the FP equation for a Dirac delta representing the initial position of a diffusing particle in an unbounded medium (the so-called propagator, Green's function or free solution) is a key quantity for the study of diffusion processes. Of course, for an arbitrary set of functions [$\ub$, $A$, $D$] no exact solution is available. However, for the important case of  the expansion $y_i=a_i(t) x_i$, the FP equation \eqref{FKfinal} can be simplified to a large extent when $A_i$ and $D_{ij}=D_i \delta_{ij}$ are constant. In this case, an exact analytical form for the propagator can be found.  This type of expansion has been extensively considered in the literature describing tissue growth \cite{Crampin1999,Crampin2001,Crampin2002,Chisholm2010,Baker2010}. In the context of Cosmology, it corresponds to a Friedmann-Lema\^{i}tre-Robertson-Walker universe where $a_i(t)=a(t)$ stands for the (Robertson-Walker) scale factor and  $H=\dot a/a$ is the so-called Hubble  parameter.  For example, for a  matter-dominated flat universe one has $a(t)\propto t^{2/3}$, whereas for a dark energy-dominated flat universe $a(t)$ grows exponentially.

We begin by considering the one-dimensional FP equation for the case of non-zero intrinsic drift, $A(\yb,t)\equiv v(\yb,t)\neq 0$ and uniform expansion:
\begin{align}
\frac{\partial }{\partial t} P(y,t)  &=
 -\frac{\partial }{\partial y} \,
\left( \frac{\dot a}{a} y+v\right)  P(y,t)
 +  D  \frac{\partial^2  }{\partial y^2 } P(y,t).
\label{FK1D}
\end{align}
Let us now define the function $P_x(x,t)=P(y=a(t) x,t)$. Then,
\begin{equation}
\frac{\partial P_x}{\partial t}=\frac{\partial P}{\partial t }+\frac{\dot a }{a} y\,  \frac{\partial P }
{\partial y}.
\end{equation}
Inserting this result into Eq.~\eqref{FK1D} and performing the transformation $y=a(t) x$  one finds
\begin{align}
\frac{\partial }{\partial t} P_x(x,t)  &=
 -\frac{\dot a}{a}\,  P_x(x,t)
 -\frac{1}{a} \frac{\partial}{\partial x}  v P_x(x,t)
 +  \frac{D }{a^2} \frac{\partial^2  }{\partial x^2 }      P_x(x,t).
  \label{FK1Da}
\end{align}
Next, we perform the change of variables $P_x(x,t)=Q(x,t)/a(t)$ in the previous equation and obtain
\begin{align}
\frac{\partial }{\partial t} Q(x,t)  &=
-\frac{1}{a} \frac{\partial }{\partial x} v\, Q(x,t)
  + \frac{D}{a^2} \frac{\partial^2  }{\partial x^2 } Q(x,t).
  \label{FK1Db}
\end{align}
Let us now focus on the case where the intrinsic drift is absent ($v=0$). We perform the following time scale transformation
\begin{equation}
\label{taudef}
\tau(t)=\int_{t_0}^t \frac{ds}{a^2(s)}.
\end{equation}
In what follows, we shall often refer to $\tau$ as ``the Brownian conformal time'' by analogy with the (standard or ballistic) conformal time defined by the equation $\dot \tau_c=1/a$ in the context of Cosmology.  In terms of the Brownian conformal time $\tau$ and the comoving coordinate $x$, Eq.~\eqref{FK1Db} becomes identical with the standard diffusion equation
\begin{align}
\frac{\partial }{\partial \tau} Q(x,\tau)  &=
  D  \frac{\partial^2  }{\partial x^2 } Q(x,\tau).
  \label{FK1Dc}
\end{align}
The solutions of Eq.~\eqref{FK1Dc} are well-known.  The propagator in physical space can be obtained by means of the inverse transformation $P(y,t)= Q[y/a(t),\tau(t)]/a(t)$.

We are now in the position to easily obtain the propagator for the case of a uniform expansion with no intrinsic drift, i.e., the solution of Eq.~\eqref{FK1D} with $A=0$ and the initial condition $P(y,t_0)=\delta(y)$.  Recall that $a(t_0)=1$ and $\tau(t_0)=0$, implying that the initial condition for the $Q$ function is $Q(x,0)=\delta(x)$.  For this initial condition, the solution of Eq.~\eqref{FK1Dc} corresponding to an unbounded system is the well-known Gaussian function:
\begin{equation}
Q(x,\tau)= \frac{1}{\sqrt{4\pi D \tau} }\, e^{-x^2/4D\tau}\equiv Q_G(x,t;D)\equiv Q_G(x,t).
\end{equation}
For $v=0$, the propagator $G(y,t)$ for diffusion in a uniformly expanding medium is then given by
\begin{equation}
\label{PytPropa}
G(y,t)= \frac{1}{a(t)}\, Q_G\left[\frac{y}{a(t)},\tau(t)\right]=
\frac{1}{\sqrt{4\pi D a^2(t)\tau}  }\, e^{-y^2/4Da^2(t)\tau }.
\end{equation}

For non-zero drift $v\neq 0$, the moments $\langle y^m\rangle=\int_{-\infty}^\infty G(y,t) y^m dy$ of the walker's position can be obtained directly by multiplying  Eq.~\eqref{FK1D} with $y^m$ and by integrating the resulting equation over $y$. Subsequent application of partial integration finally yields the following equation for the first-order moment:
\begin{equation}
\frac{d\langle y \rangle}{dt}  = \frac{\dot a}{a}\, \langle y \rangle +\langle v \rangle,
\end{equation}
whose solution is
\begin{equation}
\label{yavgt}
\langle y \rangle =a(t)\int_{t_0}^t \frac{\langle v \rangle}{a(s)} \,ds.
\end{equation}
This is precisely the proper distance traveled by a particle with velocity $\langle v \rangle$ during the time interval $[t_0,t]$ \cite{Ryden2003}. The equation for the second moment of $y$ is
\begin{equation}
\label{y2aveq}
\frac{d}{dt} \langle y^2 \rangle =2 \frac{\dot a}{a}\, \langle y^2 \rangle + 2  \langle v\,y \rangle + 2D.
\end{equation}
When $v$ and $y$ are uncorrelated, that is, when $\langle v\,y \rangle=\langle v\rangle \langle y \rangle$, one can insert Eq.~\eqref{yavgt} into  Eq.~\eqref{y2aveq} to obtain a first-order equation with a single unknown, namely, $\langle y^2 \rangle$.  For $v=0$ the solution is quite simple: $\langle y^2 \rangle=2D \, a^2(t)\,\tau(t)$ or, in terms of the comoving distance, $\langle x^2 \rangle=2D \,\tau(t)$.
We can use these expressions to rewrite Eq.~\eqref{PytPropa} in a especially simple way:
\begin{equation}
\label{PytPropaSimple}
G(y,t)=
\frac{1}{\sqrt{2\pi \langle y^2\rangle}}\, e^{-y^2/2\langle y^2\rangle }.
\end{equation}
Finally, it should be noted that the above procedure can be easily extended to derive the full hierarchy of moments, i.e.,
\begin{equation}
\frac{d}{dt} \langle y^m \rangle =m \frac{\dot a}{a}\, \langle y^m\rangle + m  \langle v \, y^{m-1} \rangle + m(m-1) D\langle y^{m-2} \rangle.
\end{equation}

For the case with intrinsic drift, $v\neq 0$, the propagator is
\begin{equation}
 G(y,t)=\frac{1}{a(t)} Q_G\left[\frac{y-\langle y\rangle}{a(t)},\tau(t)\right],
\end{equation}
with $\langle y\rangle$ given by Eq.~\eqref{yavgt}, as can be checked by inserting the above expression into Eq.~\eqref{FK1D}.

The generalization of  the above results  to  $d$-dimensional systems with $\ub=\{a_i(t) x_i\}$, $\vb= \{ v_i(y_i) \} $, and  $D_{i,j}=D_i \delta_{i,j}$ is immediate.  For example, the components of the first moment are
\begin{equation}
\label{yiavgt}
\langle y_i \rangle =a_i(t)\int_{t_0}^t \frac{\langle v_i \rangle}{a_i(s)} ds,
\end{equation}
whereas the second moment is $\langle \yb^2\rangle=\sum_{i=1}^d \langle y_i^2\rangle$. In this case, the propagator reads as
\begin{equation}
\label{dProp}
G(\yb,t)=\prod_{i=1}^d \frac{1}{a_i(t)} Q_{G\,i}\left[\frac{y_i-\langle y_i\rangle}{a_i(t)},\tau_i(t);D_i\right].
\end{equation}
Since in the higher dimensional case the computation of the moments from the $d=1$ moments is straightforward (especially in the isotropic case), we shall in general only give one-dimensional results in what follows.

Finally, let us note that for a general initial condition $P({\bf y},t_0)$ the solution  is given by the convolution with the Green's function $G(\yb,t)$:
\begin{equation}
P(\yb,t)=\int P({\bf y-\zb},t_0) G(\zb,t) d\zb \equiv [P(\cdot,t_0) \ast G(\cdot,t)](\yb).
\end{equation}
In particular, for $P(y,t_0)=\frac{1}{2} [\delta(y-y_0)+\delta(y+y_0)]$ one has
$P(y,t)=\frac{1}{2} [G(y-y_0,t_0)+G(y+y_0,t_0)]$.
We shall make use of this expression in Sec.~\ref{secPowerlawexp}.

\subsection{Diffusive pulses in expanding media}
\label{secDiffPulse}

It is instructive to bring out the similarities of some of the previous results with others found in Cosmology.

For a uniform medium ($a_i=a$) and random walkers with $D_i=D$  and zero drift, the propagator given by Eq.~\eqref{dProp} becomes
\begin{equation}
\label{PrtPropa}
G(\yb,t)= \frac{1}{\left[4\pi D a^2(t)\tau\right]^{d/2} }\, e^{-r^2/4D\tau },
\end{equation}
where the comoving radial distance $r=|\xb|=|\yb|/a(t)$ has been introduced. Equation~\eqref{PrtPropa}   describes the spread of a diffusive (or Brownian) pulse starting as a point source at position $\yb=0$ at time $t_0$. The standard deviation associated with such a diffusive pulse, namely, $\bar y\equiv \langle \yb^2 \rangle^{1/2}=a(t)\bar r$ with $\bar r^2=\langle \xb^2 \rangle$ and $\bar r=[2dD\tau(t)]^{1/2}$, is a measure of how far it has typically traveled after a given time $t-t_0$. Then, by analogy with the definition of the light cone in Cosmology, we can define a diffusive paraboloid of revolution made up by the points situated at a comoving distance $\le \bar r(t)$ from the initial location of the delta peak.
The paraboloid is obtained by revolving a parabola defined by the value of the distance $\bar r(\tau)$ around
the $\tau$ axis (the conformal time $\tau$ goes from $0$ to $\tau(\infty)$). Note that in two spatial dimensions the transversal section
of such a paraboloid is a circle (embedded in a plane defined by a fixed value of $\tau$), whereas in three dimensions it is a sphere,
and in higher dimensions a hypersphere.

On the other hand, the probability $\bar p$ that a walker has traveled a distance $\le \bar r$ during the time interval $t-t_0$
is simply
\begin{equation}
\label{xp}
\bar p\equiv\int_0^{\bar y} G(\yb,t) d\yb
= \frac{1}{\Gamma(d/2)}\int_0^{\bar r^2/4D\tau} u^{d/2-1} e^{-u} du
=1-\frac{\Gamma\left(d/2,d/2\right)}{\Gamma\left(d/2\right)},
\end{equation}
where $\Gamma(\cdot,\cdot)$ and $\Gamma(\cdot)$ are the incomplete and complete gamma function, respectively. To obtain the right hand side, we have taken into account that the surface of a hypersphere of radius $R$ is $s_d(R) \equiv 2\pi^{d/2}R^{d-1}/\Gamma\left(d/2\right)$, as well as the equality $\bar r^2/4D\tau=d/2$.  In view of the above, one can alternatively define the diffusive paraboloid as the locus of the points inside a $d$-dimensional hypersphere centered at $\yb=0$ whose radius is such that it contains an average fraction $\bar p$  of a collection of random walkers located at $\yb=0$ at time $t_0$.  In particular, $\bar p\approx 0.6827$ for $d=1$, $\bar p\approx 0.6321$ for $d=2$ and $\bar p\approx 0.6084$ for $d=3$.

Returning to the analogy with the definition of the light cone in Cosmology, we now introduce the characteristic distance
$\bar r_\text{Bh}=[2dD\tau(t)]^{1/2}$, which is simply the radius of the $d$-dimensional hypersphere defined in the previous
paragraph. We shall term this distance ``Brownian horizon'' at time $t$ [loosely speaking, we can say that the pulse (typically) reaches a distance $r_\text{Bh}$ at time $t$].   This definition can be compared with the usual cosmological definition of particle horizon $\bar r_\text{h}=c\tau_\text{c}(t)$ as the distance of the most distant object which can be seen at time $t$ ($c$ stands for the speed of light).
This comparison makes it natural to define the Brownian event horizon $r_\text{Beh}$ as the largest comoving distance a diffusive pulse emitted at $t_0$ typically ever reaches, that is,
\begin{equation}
\label{rBeh}
r^2_\text{Beh}=2dD\tau(\infty).
\end{equation}

As an aside, we note that the idea embodied by the formula~\eqref{xp} can be easily extrapolated to the case of a fixed comoving distance by considering the probability $p^*(t)$ that at time $t>t_0$ our Brownian particle is found inside an expanding hypersphere centered at the origin.  Let us denote by $R_0$ the radius of the hypersphere at time $t_0$. At later times $t$, the radius is given by $a(t) R_0$, and one has
\begin{equation}
p^*(t)\equiv\int_0^{a(t) R_0} G(\yb,t) d\yb
=\frac{1}{\Gamma(d/2)}\int_0^{\bar R_0^2/4D\tau} u^{d/2-1} e^{-u} du
=1-\frac{\Gamma\left[d/2,R_0^2/(4 D \tau)\right]}{\Gamma\left(d/2\right)},
\label{past}
\end{equation}
whereby the dependence on the scale factor enters the above equation solely via the conformal time $\tau(t)$.
In particular,  $p^*(t)\to 0$ when $\tau(t)\to \infty$ for $t\to\infty$ [recall that $\Gamma(\cdot,0)\equiv \Gamma(\cdot)$], while $p^*(\infty)>0$  when $\tau(t)$ remains finite at all times $t$.   For example, for a power law expansion, $a(t)=(t/t_0)^\gamma$ one finds that $p^*(t)\to 0$ when $0\le\gamma\le 1/2$, while $p^*(\infty)>0$  when  $\gamma>1/2$ [see Eqs.~\eqref{taua} and~\eqref{taub} below].
In passing, we note that this behavior  is somewhat reminiscent of the extinction phenomenon in Galton-Watson processes~\cite{grimmet}.

Equation~\eqref{past} admits a simple probabilistic interpretation, namely, $p^*(t)=\text{Pr}[\chi \le R_0^2/(4 D \tau)]$; in other words, $p^*(t)$ is equal to the probability that $\chi \le R_0^2/(4 D \tau)$,  where $\chi$ is a random variable drawn from the gamma distribution $\Gamma(d/2,1)$. When $d$ is even the following alternative probabilistic interpretation in terms of the Poisson distribution holds:
$p^*(t)=\text{Pr}[\chi \ge d/2]$, where $\chi$ is a random variable following a Poisson distribution with parameter $R_0^2/(4 D \tau)$, i.e.,  $\chi \sim \text{Poi}[R_0^2/(4 D \tau)]$.

Note that  $p^*(t)$ is an upper bound  for the so-called survival probability $\Pi(\tau(t))$, i.e., the probability that the Brownian particle never leaves the interior of the hypersphere. This is the case because the definition of $p^*(t)$ does not preclude recrossing of the boundary of the hypersphere. In Sec.~\ref{secSurvival} we show how to compute $\Pi(\tau(t))$.

\subsection{Power-law expansion}
\label{secPowerlawexp}
Let us consider in detail the case of a uniform expansion whose time evolution is described by a power law, $a(t)=(t/t_0)^\gamma$ (with $\gamma>0$).  This case is relevant for at least two reasons. The first one is that this type of expansion corresponds to a flat  Friedmann-Lema\^{i}tre-Robertson-Walker universe. For example, $a(t)\propto t^{1/2}$ corresponds to a radiation-dominated universe, while $a(t)\propto t^{2/3}$ describes the expansion of a matter-dominated universe.
The second reason is that the typical spread length of diffusive particles in a static medium only grows as $t^{1/2}$. This results in a nontrivial, interesting interplay between the two coexisting transport mechanisms, i.e., diffusion (possibly with an intrinsic bias $v$) and the drift due to the medium expansion.

For $a(t)=(t/t_0)^\gamma$ one finds
\begin{align}
\label{taua}
\tau(t)=&t_0^{2\gamma}\, \dfrac{t^{-2\gamma+1}-t_0^{-2\gamma+1}}{1-2\gamma}, \quad \gamma\neq 1/2,\\
\tau(t)=& t_0 \ln\left(\dfrac{t}{t_0}\right),\quad \gamma=1/2.
\label{taub}
\end{align}
For $v=0$ the propagator is given by Eq.~\eqref{PytPropa}, and one has  $\langle y\rangle=0$, as well as $ \langle y^2\rangle =a^2(t)\langle x^2\rangle$ with
$ \langle x^2\rangle=2D \tau(t)$, i.e.,
\begin{align}
\label{y2a}
  \langle y^2\rangle & = \frac{2Dt}{1-2\gamma} \left[ 1-\left( \frac{t}{t_0}\right)^{2\gamma-1}\right],  &\gamma\neq 1/2, \\
    \langle y^2\rangle & =  2Dt  \ln \left( \frac{t}{t_0} \right) , &\gamma= 1/2.
\label{y2b}
\end{align}

\begin{figure}[t]
\begin{center}
\includegraphics[width=0.49\textwidth,angle=0]{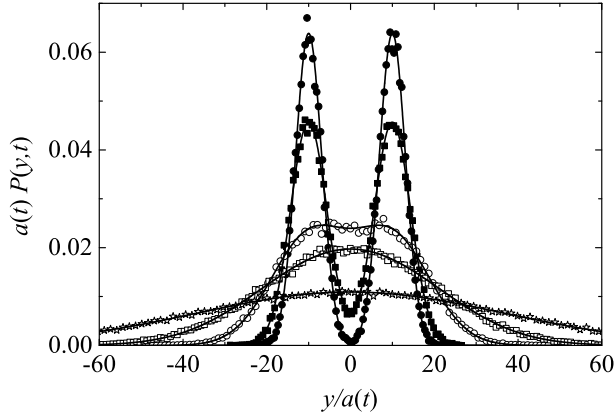}
\end{center}
\caption{\label{figPvsxga1o4} Simulation results for the probability $P(y,t)$ to find the random walker at position $y$ after a time $t$ given the initial value $P(y,0)=[\delta(y-y_0)+\delta(y+y_0)]/2$. We have set $y_0=10$, $D=1/2$ $t_0=100$ and $t-t_0=10,\, 20,\, 100,\, 500,\, 5000$ (symbols: solid circle, solid square, circle, square, and star, respectively). The medium expands according to the power law $a(t)=(t/t_0)^\gamma$ for $\gamma=1/4$. The solid lines represent the corresponding theoretical results.
}
\end{figure}

\begin{figure}[t]
\begin{center}
\includegraphics[width=0.49\textwidth,angle=0]{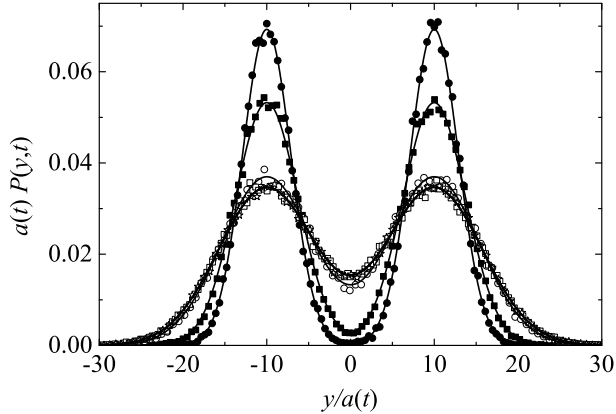}
\end{center}
\caption{\label{figPvsxga2}
Simulation results for the probability $P(y,t)$ to find the random walker at position $y$ after a time $t$ for a double-peaked initial condition  $P(y,0)=[\delta(y-y_0)+\delta(y+y_0)]/2$ representing a pair of diffusive pulses. We take $y_0=10$, $t_0=100$, $D=1/2$ and $t-t_0=10,\, 20,\, 100,\, 500,\, 5000$ (symbols: solid circle, solid square, circle, square, and star, respectively). The medium expands according to the power law $a(t)=(t/t_0)^\gamma$ and $\gamma=2$. The solid lines represent the corresponding theoretical results. There is no complete mixing between the two diffusive pulses, since the Brownian event horizon $r_\text{Beh}=[2D t_0/(2\gamma-1)]^{1/2}\approx 8.16$ is shorter than the semidistance (=10) between the two source points located at $y=\pm y_0$.
}
\end{figure}

\begin{figure}[htbp]
\centering
\subfigure[\,\,$\gamma=0$]{\includegraphics[width=.4\linewidth]{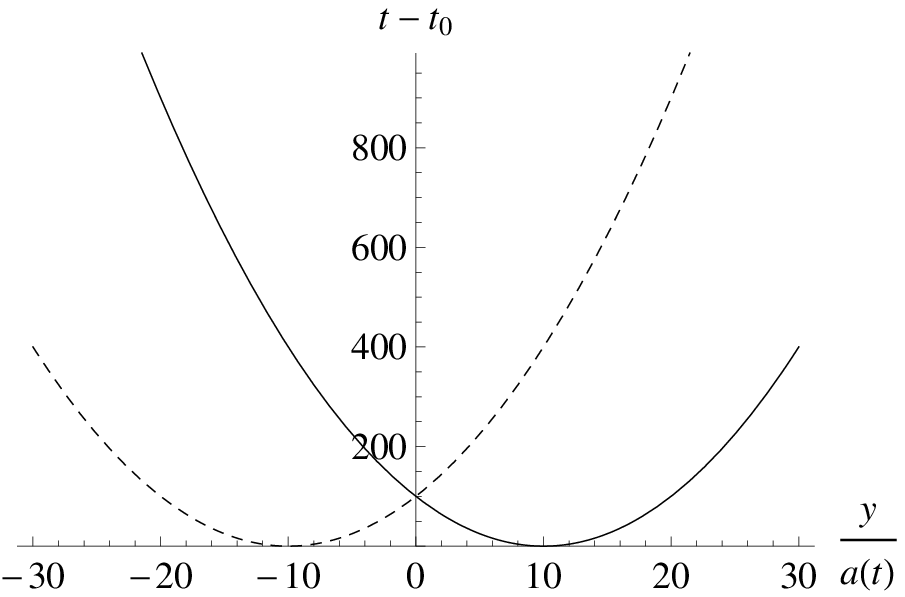}} \hspace{.4cm}
\subfigure[\,\,$\gamma=1/4$]{\includegraphics[width=.4\linewidth]{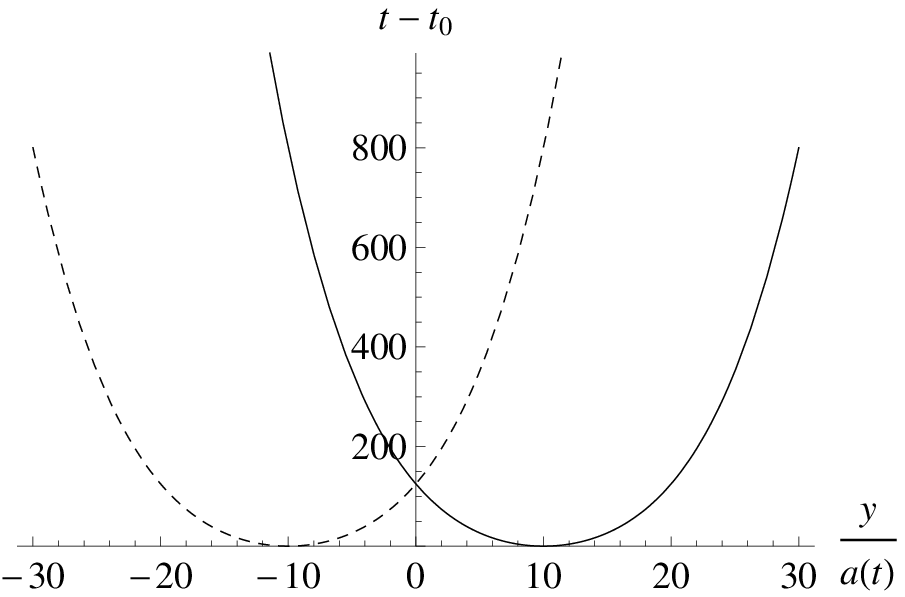}} \\ \vspace{.4cm}
\subfigure[\,\,$\gamma=1/2$]{\includegraphics[width=.4\linewidth]{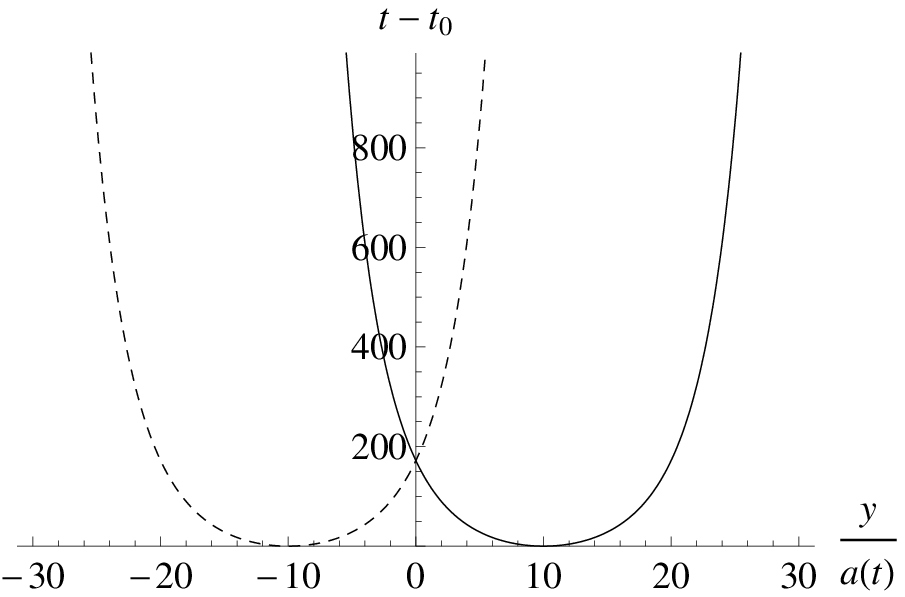}} \hspace{.4cm}
\subfigure[\,\,$\gamma=2$]{\includegraphics[width=.4\linewidth]{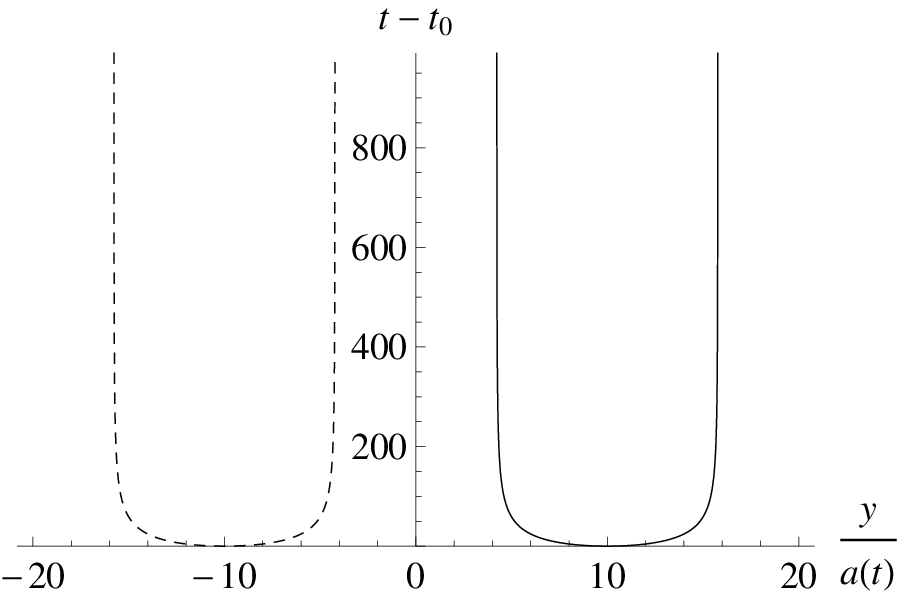}}
\caption{Brownian horizons $\bar{r}_{\text{Bh}}$ as described in the main text.
The parameter values are the same as in Figs.~\ref{figPvsxga1o4}
and~\ref{figPvsxga2} (except $\gamma$, which is specified in each subfigure).
The solid (dashed) line corresponds to the Brownian pulse
that is emitted at $y=10$ ($y=-10$) at time $t_0$. It is clear that in comoving coordinates both pulses become narrower as
$\gamma$ grows. Note also that for $\gamma=1/4$ both pulses overlap, but they do not for $\gamma=2$, in agreement with what is
shown in Figs.~\ref{figPvsxga1o4} and \ref{figPvsxga2}.}
\label{horizonfig}
\end{figure}

Note that for $\gamma<\frac{1}{2}$ and long times the growth of the typical spread length, $\langle y^2\rangle^{1/2}\propto t^{1/2}$  is faster than the domain expansion, $a(t)\propto t^\gamma$,  implying that the diffusing particles spread across the full expanding domain.
In contrast, for $\gamma>\frac{1}{2}$ and long times  the diffusive spread length grows as fast as the scale factor [$\langle y^2\rangle^{1/2}\propto t^\gamma$  vs. $a(t)\propto t^\gamma$]. This means that the medium expansion dominates, and particles are not able to efficiently spread across the medium, resulting in strong localization effects. An alternative way to see this consists in examining the long-time behavior of the second moment of the traveled distance expressed in comoving coordinates, $\langle y^2 \rangle/a^2(t)=\langle x^2 \rangle= 2 D \tau(t)$. From Eqs.~\eqref{taua}-\eqref{taub} one finds
\begin{equation}\label{eqcm2}
\langle x^2 \rangle \sim \begin{cases}
2D \frac{t_0^{2\gamma}}{1-2\gamma}t^{1-2\gamma},& \quad  \gamma < \frac{1}{2}, \\
2D t_0 \ln\left(t\right),& \quad \gamma = \frac{1}{2}, \\
2D \frac{t_0}{2\gamma-1},& \quad   \gamma > \frac{1}{2},
\end{cases}
\end{equation}
in the limit $t \to \infty$.
Thus, for $\gamma<1/2$ (which includes the case of contracting media $\gamma<0$)
one has $\langle x^2 \rangle \to\infty$ as $t\to \infty$; in the language coined in Sec.~\ref{secDiffPulse} by analogy with universe expansion models used in Cosmology, the Brownian event horizon $r_\text{Beh}$ is infinite in this case,  which means that the particles are eventually able to spread across the full size of the expanding domain. This behavior is clearly seen in Fig.~\ref{figPvsxga1o4}, where two diffusive pulses corresponding to two sets of non-interacting particles that initially start at $y=-y_0$ and $y=y_0$ are shown.  One sees that, after a certain time (e.g., for times $\ge 5000$), diffusional particle mixing blurs the double-peaked initial condition almost entirely.
However,  for  $\gamma> 1/2$ one has $\langle x^2 \rangle \to \text{constant}$ as $t\to \infty$,   so that  factoring out the effect of the medium expansion one sees that the spread of particles due to diffusion becomes less and less relevant in the course of time. This effect is remarkable, since it implies that for a sufficiently fast expanding medium ($\gamma>1/2$) the initial condition of the system is much less blurred in comparison with a standard diffusive process in a static medium; therefore, a remnant of this initial condition persists for arbitrarily long observation times. This is simply a consequence of the fact that the Brownian event horizon [cf. Eq.~\eqref{rBeh}] is finite in this case: $r_\text{Beh}=[2D t_0/(2\gamma-1)]^{1/2}$. Then, two diffusive pulses whose initial separation is larger than $2r_\text{Beh}$ never (effectively) meet, that is, their overlap or mixing is limited. These effects are shown in Fig.~\ref{figPvsxga2} where two diffusive pulses are again initially located at $y=-y_ 0$ and $y=y_0$. However, in the present case, the diffusional mixing is strongly hindered by the medium expansion,  and the trace of the double-peaked initial condition persists for arbitrarily long times, as opposed to the behavior shown in Fig.~\ref{figPvsxga1o4}. Note that the curves for $t=100$, 500, and 5000 are almost coincident. The evolution of the Brownian horizons for the cases corresponding to Figs.~\ref{figPvsxga1o4} and~\ref{figPvsxga2} are shown in Fig.~\ref{horizonfig}, together with the classical case ($\gamma=0$) and the marginal case ($\gamma=1/2$), which are plotted for comparison. Note the qualitative agreement of the results plotted in this last figure with those shown in the previous two figures.

The simulation results shown in Figs.~\ref{figPvsxga1o4} and \ref{figPvsxga2} were obtained from a collection of random walkers performing jumps after each time unit, whereby the jump length was drawn from a Gaussian distribution with zero mean and unit variance (this amounts to setting $D=1/2$). A total of $10^4$ runs were performed for $t-t_0=10,\, 20$,  whereas $5\times 10^4$ runs were performed for $t-t_0=100,  500,\, 5000$.

For $\gamma = \frac{1}{2}$  the domain expansion  makes the dispersion of the particles within the system only marginally faster, since the second moment $\langle y^2\rangle  =  2Dt$ is only  increased by the logarithmic factor $\ln(t/t_0)$ with respect to the case of a static medium.

For the sake of completeness, we also give below the first and the second moment of the physical distance for the case of a constant non-zero velocity drift,  $v\neq 0$:
\begin{align}
  \langle y\rangle &=\frac{vt}{1-\gamma}\left[  1- \left( \frac{t}{t_0} \right)^{\gamma-1} \right] , &\gamma\neq 1, \\
  \langle y\rangle &= vt  \log  \left(\frac{t}{t_0} \right) , &\gamma= 1.
\end{align}
The second moment $\langle y^2\rangle$ is  obtained by performing the replacement $\langle y^2\rangle\to\langle y^2\rangle -\langle y\rangle^2$ in Eqs.~\eqref{y2a} and \eqref{y2b}.

Finally, it is interesting to note that for power-law expansion and  $v=0$, Eq.~\eqref {FK1Db}  has the form of a Batchelor's equation, giving rise to the so-called scaled Brownian motion \cite{Metzler2014b,Thiel2014}. This means that, in the comoving representation, the diffusion process is non-stationary and non-ergodic in a way similar to continuous time random walks displaying memory effects~\cite{Thiel2014}.

\subsection{Exponential expansion}
\label{expexp}

Let us now consider  the case of  the exponential expansion  $a(t)= \exp[H(t-t_0)]$  with $H$ being the Hubble parameter. This case corresponds to a dark energy dominated flat universe, and it also describes the growth of many biological media, at least in the early stages~\cite{binder2008}. From the definition of Brownian conformal time, Eq.~\eqref{taudef}, one easily finds
\begin{align}
\label{tauExpo}
\tau(t)= \frac{1}{2H}\left[ 1-e^{-2H(t-t_0)}\right].
\end{align}
Then, $\langle x^2(t) \rangle = (D/H)\left[ 1-e^{-2H(t-t_0)}\right]$
and
\begin{align}
\label{y2Expo}
\langle y^2(t) \rangle = \frac{D}{H}\left[ e^{2H(t-t_0)}-1\right].
\end{align}

For $H>0$ (expanding medium), one obtains $\tau(\infty)=t_H/2$, where $t_H=1/H$ stands for the  Hubble time. The Brownian event horizon is then $r_\text{Beh}=\sqrt{2D\tau(\infty)}=\sqrt{D t_H}$, which could be termed the ``Brownian Hubble distance'', as two points separated by an initial comoving distance $x$ larger than $r_\text{Beh}$ cannot be connected by a Brownian pulse; in other words, sets of  walkers starting at points separated by $r_\text{Beh}$ cannot be effectively mixed. This is completely similar to the behavior we described in Sec.~\ref{secPowerlawexp} for the power law expansion with $\gamma>1/2$.

For $H<0$ we have a \emph{contracting} medium. In this case $\tau(\infty)$ and the  Brownian Hubble distance are infinite.  This means that, no matter how far two points are initially separated, either of the two will eventually be reached by a Brownian pulse starting from the other one. In other words, any double-peaked initial distribution of walkers will result in both peaks eventually merging into a single one. In fact, one finds from Eq.~\eqref{y2Expo} that $\langle y^2(\infty) \rangle=-D/H$, implying that the tendency of the particles to spread out due to diffusion is eventually compensated by the contracting drift of the medium. In this way,  a stationary state where the particles diffuse in a region of finite size of order $\ell_H=-D/H$ (a kind of contractive Hubble length) is reached. According to Eq.~\eqref{PytPropaSimple},  for a delta-peaked initial condition the stationary particle distribution in this region is
\begin{equation}
\label{PyEstatica}
G_s(y)=
\frac{1}{\sqrt{2\pi \ell_H^2}}\, e^{-y^2/2\ell_H^2}.
\end{equation}
For any given initial probability distribution $P_0(y)$, the resulting stationary distribution $P_s(y)$ is given by convolution $P_s(y)=[G_s \ast P_0](y)$.

\begin{figure}[t]
\begin{center}
\includegraphics[width=0.49\textwidth,angle=0]{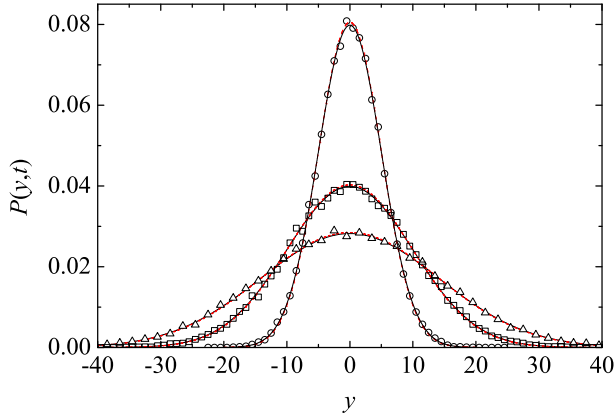}
\end{center}
\caption{\label{figPyExpo}
Simulation results for the probability $P(y,t)$ to find random walkers at position $y$ after a time $t$  in an exponential \emph{contracting} medium with $P(y,0)=\delta(y)$, $t_0=100$, $D=1/2$ and  $t-t_0=100$ for $H=-1/50$ (circles), $t-t_0=400$ for $H=-1/200$ (squares), and $t-t_0=800$ for $H=-1/400$ (triangles).  The solid lines represent the corresponding final stationary distribution $P_s(y)$ as given by Eq.~\eqref{PyEstatica}. The corresponding theoretical distributions $P(y,t)$ are also plotted (broken lines), but they are hardly distinguishable from the stationary distribution.
}
\end{figure}

In Fig.~\ref{figPyExpo} we compare this distribution for several values of $H$ with simulation results for $P(y,t)$ and long enough times ($t-t_0=100, 400, 800$ for $H=-1/50, -1/200, -1/400$, respectively), so that changes in $P(y,t)$ are barely noticeable.  As expected, these times scale as $t-t_0\propto H^{-1}$.
The simulation results were obtained from  $5\times 10^4$ random walk realizations, whereby each walker performed a jump after each time unit and the jump length was drawn from a Gaussian distribution with zero mean and unit variance.

The onset of a stationary distribution is an  exclusive feature of the case with exponential contraction. Notice, for example, that no stationary distribution is reached in the case of power-law contraction, $\gamma<0$.

\section{Survival probability and first-passage time distribution for the case of a uniform expansion}
\label{secSurvival}

So far, we have only considered the free propagator solution of the diffusion equation. However, problems with absorbing boundaries are of fundamental importance, as they provide a standard route to compute  first-passage properties. In turn, the latter are of special relevance in the context of diffusion-controlled reactions, where the corresponding reaction rates are essentially limited by the time needed to attain the reactive interface (or the interaction radius in the case of binary reactive collisions). In what follows we discuss a basic class of problems associated with an absorbing boundary condition, namely, the computation of the survival probability of particles enclosed by an expanding, fully absorbing hyperspherical surface.

Consider a Brownian point particle with $v=0$ in physical space. We assume that the particle is placed at the center of a hypersphere of expanding radius $R_y=R_y(t)=a(t) R_0$, where $R_0$ denotes the initial radius.  We ask for the probability ${\Pi}(t)$ that the particle has \emph{not} escaped from the expanding region defined by the hypersphere up to time $t$. This problem can be solved by making the surface of the hypersphere fully absorbing and by identifying the escape process (surface crossing) with absorption, which justifies the use of the term ``survival probability'' for ${\Pi}(t)$. This quantity can be obtained from the solution of the $d$-dimensional version of Eq.~\eqref{FK1D} for $v=0$, namely,
\begin{align}
\frac{\partial }{\partial t} P(\yb,t)  &=
 -\frac{\dot a}{a}\nabla\cdot [\yb P(\yb,t) ]
 +  D   \nabla^2  P(\yb,t).
  \label{FKdD}
\end{align}
The above equation must be complemented with the delta-peaked initial condition $P(r_y,t=t_0)=s_d(r_y)^{-1}\delta_+(r_y)$ [where $s_d(r_y)$ is the surface of a hypersphere of radius $r_y$] and the absorbing boundary condition $P(r_y=R_y, t)=0$. One also has the implicit condition that $P(r_y, t)$ must remain finite everywhere at all times.
The notation $\delta_+(\cdot)$ has been used for the slightly modified delta-function with the property $\int_0^R \delta_+(r) dr=1$ for any $R>0$. Once the corresponding solution $P(r_y, t)$  is known, the survival probability follows immediately as $\Pi(t)=\int_0^{R_y} P(r_y, t)\, s_d(r_y) dr_y$.

Proceeding as in Sec.~\ref{secGruex}, it is possible to reduce Eq.~\eqref{FKdD} to a simpler form by introducing comoving coordinates $|\xb|\equiv r_x= r_y/a(t)$ as well as a new function $Q$ defined by the substitution $P(r_y,t)= Q[r_x,\tau(t)]/a(t)$. The resulting equation, $\partial Q/\partial \tau=D\nabla^2Q$, is the $d$-dimensional generalization of Eq.~\eqref{FK1Dc} for the case of a hyperspherical geometry, i.e.,
\be
\frac{\partial Q(r, \tau)}{\partial \tau}=D\;\left\{\frac{\partial^2 }{\partial r^2}+\frac{d-1}{r}\frac{\partial}{\partial r}
 \right\} Q(r,\tau).
\label{CMFPEtau}
\ee
In the above equation we have set $r_x\equiv r$; this notation will be used throughout the remainder of the present section. Taking into account the equations $\tau(t_0)=0$ and $a(t_0)=1$ as well as the initial and boundary conditions for $P(r_y,t)$, one finds
$Q(r,\tau=0)=s_d(r)^{-1}\delta_+(r)$  and $Q(r=R_0, \tau)=0$.
In addition, the normalization condition $\int_0^{R_0} P(r, t_0)\,s_d(r) dr=\int_0^{R_0}
Q(r, \tau=0)s_d(r)\,dr \equiv 1$ must be fulfilled. The well-known solution to the above problem (easily found by separation of variables) can be written as follows \cite{Borrego2009,Yuste2010}:
\be
Q(r,\tau)=
\sum_{n=1}^{\infty} \left(\frac{j_n}{2R_0}\right)^{d/2-1} \frac{r^{1-d/2}
}{\pi^{d/2} R_0^2 J_{d/2}^2(j_n)} J_{d/2-1}\left(
\frac{j_n\,r}{R_0}\right) \,e^{-j_n^2 D \tau/R_0^2},
\label{propagator2}
\ee
where $j_n\equiv j_{d/2-1, n} $ is the $n$-th positive zero of the Bessel function of order $d/2-1$, i.e., $J_{d/2-1}(j_{d/2-1, n})=0$. For simplicity, we use the short-hand notation $j_n$; however, the reader should bear in mind that $j_n$ depends on the spatial dimension $d$.

In terms of $Q$, the survival probability $ {\Pi}(t)= \Pi[\tau(t)]$ is expressed as
\begin{equation}
 \Pi(\tau)=\int_0^{R_0} Q(r,\tau) s_d(r)\, dr =
 \frac{2^{2-d/2}}{\Gamma(d/2)}
\sum_{n=1}^{\infty}\frac{j_n^{d/2-2}}
{J_{d/2}(j_n)} e^{-j_n^2 D \tau/R_0^2}.
\label{PiTauExp}
\end{equation}
We see that the behavior of $\Pi(t)$ depends on how $\tau(t)$ behaves.  For example, if $\tau(t\to\infty)\equiv\tau_\infty\neq 0$, then the probability that a particle is never trapped is simply $\Pi(\tau_\infty)$, a non-zero quantity.  This is the case for the previously defined power-law expansion with $\gamma>1/2$, which gives $\tau_\infty=t_0/(2\gamma-1)$ [cf. Eq.~\eqref{taua}]. On the other hand, if $\tau_\infty=\infty$, then the probability  $\Pi(\tau_\infty)$ that a particle is never trapped is zero.  This is the case, for example, for the power-law expansion with $\gamma\leq 1/2$. A previous derivation of the above results for $d=1,2,3$ has been given in Refs.~\cite{Simpson2015b, Simpson2015}.

There is an alternative and instructive way to see that $\Pi(\tau_\infty)$ must vanish when $\tau_\infty=\infty$. In this case, it is possible to define the Laplace transform of the survival probability as
\begin{equation}
  \tilde{\Pi}(u)\equiv \int_0^\infty e^{-u\tau}\Pi(\tau)\,d\tau,
 \label{lapsurvprob}
\end{equation}
since the conformal time variable $\tau$ spans the full range of positive real numbers.
It turns out that the analytic form of the Laplace transform given by Eq.~\eqref{lapsurvprob} is known
\cite{redner}:
\be
\tilde{\Pi}(u)=\frac{1}{u}-\frac{2^{1-d/2}}{u}
\frac{\left(u R_0^2/D\right)^{(d/2-1)/2}}
{\Gamma(d/2)I_{d/2-1}(\sqrt{\frac{u}{D}}R_0)}.
\label{LTSPb}
\ee
Hence, the final value theorem yields
\be
\lim_{t\to\infty} \Pi[\tau(t)]
=\lim_{\tau\to\infty} \Pi(\tau)=\lim_{u\to 0} u\,\tilde{\Pi}(u) =0
\ee
regardless of the value of the spatial dimension.

Finally, we note that the result expressed by Eq.~\eqref{propagator2} for a delta-peaked initial condition is just a particular case of the general problem with  the (hyperspherical) initial condition  $P(r_y, t_0)\equiv Q(r,0)$, as described in, e.g., Ref.~\cite{Carslaw1959} or in  Refs.~\cite{Borrego2009,Yuste2010} for the case of a subdiffusive particle (in this last case, one must replace the Mittag-Leffler functions appearing in those references with exponential functions). Thus, one obtains
\be
Q(r,\tau)=\sum_{n=1}^{\infty} a_n \, r^{1-d/2} J_{d/2-1}
\left(\frac{j_n\,r}{R_0}\right) \,e^{-j_n^2 D \tau/R_0^2}
\ee
with
\be
a_n=\frac{2}{R_0^2 J_{d/2-1}^2 \left( z_n \right)}\int_0^{R_0} r^{d/2} Q(r,0)J_{d/2-1}\left(\frac{j_n\,r}{R_0}\right)\,dr.
\ee
In dimensions $d=1$ and $d=3$ the aforementioned initial condition leads to coefficients $a_n$ which are expressible in terms of trigonometric functions, whereas in $d=2$ the corresponding $a_n$'s are given by Bessel functions.
In Ref.~\cite{Simpson2015}, a specific form of the initial condition was studied, namely,
$Q(r,0) \propto 1-\Theta(r-r_0)$, where $\Theta(\cdot)$ stands for the Heaviside step function, and $r_0 \le R_0$.
This kind of initial condition is relevant for the development of the enteric nervous system ~\cite{Landman}.

\subsection{Moments of the first-passage time for power-law expansion}

The moments of the first-passage time $\langle t^n \rangle$ can be straightforwardly computed from the first-passage time distribution $F(t)= -d\Pi(t)/dt$. The $m$-th-order moment is given by the following formula:
\be
\langle t^m \rangle = \int_{t_0}^\infty dt \, F(t) t^m, \qquad m=0, 1, 2, \ldots
\label{m-moment}
\ee
or, integrating by parts,
\be
\langle t^m \rangle =  -\left. t^m \Pi(t) \right|_{t_0}^\infty
+m\int_{t_0}^\infty t^{m-1}\Pi(t) dt.
\label{FPTmomentsGral}
\ee
We shall consider different subcases for the power-law scale factor $a(t)=(t/t_0)^\gamma$ depending on the value of $\gamma$.

\subsubsection{Case $\gamma<1/2$}

In this case Eq.~\eqref{PiTauExp} gives
\be
\Pi(t)=\frac{2^{2-d/2}}{\Gamma(d/2)}
\sum_{n=1}^{\infty}\frac{j_n^{d/2-2}}
{J_{d/2}(j_n)} \mbox{exp}\left[{-j_n^2 D t_0^{2\gamma} \frac{t^{1-2\gamma}-t_0^{1-2\gamma}}{(1-2\gamma) R_0^2}}\right].
\ee
This can be expressed in a more compact way as follows:
\be
\Pi(t)=\sum_{n=1}^\infty  \rho_n e^{-\alpha_n (t^{1-2\gamma}-t_0^{1-2\gamma})},
\label{rhoprimeexp}
\ee
where the quantities $\rho_n= 2^{2-d/2}\,j_n^{d/2-2}/[\Gamma(d/2)\,J_{d/2}(j_n)]$
and $\alpha_n=j_n^2D R_0^{-2} t_0^{2\gamma}/(1-2\gamma)$ have been introduced. For $t=t_0$ the series $\Pi(t_0)=\sum_{n=1}^\infty\rho_n$ is divergent for $d\ge 3$.
This singularity in the initial condition is well-known from the analogous diffusion problem in static domains; however, we know that the physical value of the sum is $\sum_{n=1}^\infty\rho_n \equiv 1$ in all dimensions. On the other hand, this value can be recovered by regularizing this divergent series. To this end, a technique akin to  Abel summation \cite{Hardy} can be applied, whereby suitable regulator functions involving Bessel functions are employed \cite{Yuste2010}. In particular,  $\sum_{n=1}^\infty\rho_n$ is just $2^{2-d/2}/\Gamma(d/2)$ times the series denoted by $S(d/2-1,0)$ in Ref.~\cite{Yuste2010}, which is equal to $2^{d/2-2}\Gamma(d/2)$ [see the result below Eq.~(26) in Ref.~\cite{Yuste2010}].

When $\gamma <1/2$, $\Pi(t)\to 0$ for $t\to\infty$ and Eq.~\eqref{FPTmomentsGral} becomes
\be
\langle t^m \rangle =t_0^m+m\int_{t_0}^\infty t^{m-1}\Pi(t) dt
\label{FPTmoments}
\ee
as  $\Pi(t_0)\equiv 1$ by construction. Taking Eq.~\eqref{rhoprimeexp} into account, one obtains
\be
\langle t^m \rangle =t_0^m+m\sum_{n=1}^\infty  \rho_n e^{\alpha_n t_0^{1-2\gamma}}\int_{t_0}^\infty t^{m-1} e^{-\alpha_n t^{1-2\gamma}} dt.
\ee
The above expression for $\langle t^m \rangle$ can be rewritten in terms of incomplete Gamma functions. One has
\be
\langle t^m \rangle=t_0^m+\frac{m}{1-2\gamma}\sum_{n=1}^\infty \rho_n
\alpha_n^{-\frac{m}{1-2\gamma}} e^{\alpha_n t_0^{1-2\gamma}}\,\Gamma\left(\frac{m}{1-2\gamma}, \alpha_n t_0^{1-2\gamma}\right).
\label{seriestm}
\ee
Note that, since we have assumed $\gamma<1/2$, one has $\alpha_n>0$. On the other hand, for a fixed value of the spatial dimension one has $j_n \to [n+(d-3)/4]\pi$ for large $n$ according to McMahon's asymptotic expansion \cite{Abra}. Besides, for fixed order and large values of the argument, the following asymptotic expansion of the Bessel function holds
\cite{Abra}:
\be
J_\nu(z) \sim \sqrt{\frac{2}{\pi z}} \mbox{cos}\left(z-\frac{\nu \pi}{2}-\frac{\pi}{4}\right), \qquad
|z|\to\infty,
\ee
implying that $J_{d/2}(j_n)\to (-1)^{n-1}[2/(\pi^2 n)]^{1/2}$ for $n \to \infty$. Using the large-$x$ approximation $\Gamma(a, x)\sim x^{a-1} e^{-x}$ one finds that the series expansion~\eqref{seriestm} converges for arbitrary $m>0$ in one, two and three dimensions.

\subsubsection{Case $\gamma=1/2$}

Taking $\tau=t_0 \mbox{ln}(t/t_0)$ in Eq. \eqref{PiTauExp} and using the definition of $\rho_n$ we obtain
\be
\Pi(t)=\sum_{n=1}^\infty \rho_n \left(\frac{t_0}{t}\right)^{\eta_n}
\ee
with $\eta_n\equiv j_n^2 D t_0/R_0^2$. For $t>t_0$ this series tends to zero as $t\to\infty$ in any spatial dimension. Let us now examine the behavior of the moments of the first-passage time. In this case Eq.~\eqref{FPTmomentsGral} becomes
\be
\langle t^m\rangle = -\left. \sum_{n=1}^\infty \left( 1+\frac{m}{\eta_n-m}\right) \rho_n t_0^{\eta_n}\, t^{m-\eta_n} \right|_{t=t_0}^{t=\infty}
\ee
when $m\neq \eta_n$. Since the $\eta_n$'s increase monotonically with $n$, it is necessary and sufficient that $\eta_1=j_1 D t_0/R_0^2>m$ for $\langle t^m\rangle$  to be finite. When this is the case, the upper boundary term vanishes and one finally obtains
\be
\langle t^m \rangle= t_0^m+m \sum_{n=1}^\infty \rho_n \frac{t_0^m}{\eta_n-m}.
\label{etaseries}
\ee
For large $n$ one has $\eta_n\propto j_n\propto n$, and $\rho_n \propto (-1)^{n-1} n^{(d-3)/2}$. Hence the above series converges in one, two and three dimensions for any $m>0$. The condition $\eta_1>1$ implies  that the diffusion coefficient must exceed a threshold value,
\be
D> \frac{1}{j_{d/2-1,1}} \frac{R_0^2}{t_0},
\ee
for the mean first-passage time $\langle t\rangle$  to exist. We have restored the full notation for the $j_n$'s to emphasize the dependence on dimensionality. In more general terms, if the condition $m+1\ge \eta_1>m$ holds, i.e., if
\be
\frac{m+1}{j_{d/2-1,1}} \frac{R_0^2}{t_0}\ge D>\frac{m}{j_{d/2-1,1}} \frac{R_0^2}{t_0}
\ee
holds, the $m$-th moment of the diffusion coefficient is still finite, but neither the $m+1$-th moment nor higher order moments exist in one, two or three dimensions.  Note that when $\eta_1=m$ the $m$-th moment diverges logarithmically.

\subsubsection{Case $\gamma>1/2$}

As already mentioned, Eq. \eqref{PiTauExp} also holds in this case and one finds $\Pi(\tau_\infty)>0$ with $\tau_\infty=t_0/(2\gamma-1)$. Thus, since $\Pi(t\to\infty)\neq 0$, neither the mean first-passage time nor higher order moments exist.

\section{Concluding remarks}
\label{conclusions}

In this work, a Chapman-Kolmogorov equation for diffusion in an expanding medium has been obtained and subsequently employed to deduce the corresponding $d$-dimensional FP equation. The free solution or propagator in physical space $P(\yb,t)$ has been explicitly obtained for the case of uniform expansion. Typical properties associated with the diffusive spread of particles in expanding media have been investigated in terms of what we call Brownian  horizon, a characteristic distance somewhat analogous to the particle horizon defined in Cosmology.

We have subsequently focused our discussion on the important case of a uniform expansion with power-law scale factor $a(t)\propto t^\gamma$. The value $\gamma=1/2$ plays a special role, as it separates the regime of complete mixing (infinite Brownian event horizon, $\gamma<1/2$) from the regime of truncated or imperfect mixing (finite Brownian event horizon, $\gamma>1/2$). Theoretical results for the probability distribution functions in these two regimes have been confirmed by means of numerical simulations.

Finally, we have considered diffusion problems in the presence of a fully absorbing hyperspherical boundary. We have confirmed and extended previous results for the survival probabilities of a particle initially localized at the center of an expanding hypersphere with a fully absorbing surface. In this case, we have discussed interesting crossover effects in the context of a uniform medium expansion described by a power-law, both at the level of the survival probability and of the moments of the first-passage time distribution.

We see this work as a step towards a stochastic theory of diffusion in expanding spaces. As discussed in the Introduction, our motivation was originally fueled by important problems in connection with Cosmology and Biology, but we anticipate that a variety of other systems where the medium expansion occurs on time scales commensurate with diffusive transport are likely to display similar features.

Regarding possible extensions of the present work, we favor two main lines of research. The first one is rather fundamental in nature, as it aims to enlarge the theoretical framework for non-equilibrium statistical mechanics in expanding spaces by considering purely diffusive systems as a first step towards a more general description of a wide class of reaction-diffusion systems.
 In this context, an interesting example concerning a coupled set of 1d linear reaction-diffusion equations describing cell proliferation within a growing tissue has recently been studied in Ref.~\cite{Simpson2015c}. It would be interesting to extend this study by allowing for a time dependence of the rate constants in the linear reaction terms, in which case it should be possible to find an uncoupling transformation similar to the one used in this reference. The extension to higher dimensions or different types of boundary conditions appears to be less straightforward, but it is also of interest.

A second line of research addresses the connection of our theory with experiments and astrophysical observations. Of particular interest in this context is the fact that, for a radiation-dominated universe one has $a(t) \propto t^{1/2}$. According to our findings, this is precisely the critical expansion rate separating the regime in which the walker remains strongly localized at all times (and where a strong memory of the initial condition persists) from a regime where the walker delocalizes rather quickly (essentially as it happens in a static medium).  In view of this, it is possible that non-trivial probabilistic effects happened in some stages of the Universe evolution. On the other hand, given the wide plethora of behaviors found in Biology, it would be surprising that such effects were not present in certain types of biological systems too.
In this context, we wish to point out that the case of a medium whose expansion saturates in the course of time  may be relevant for the phenomenology of certain living systems~\cite{binder2008}.  In this case, the scale factor may be logistic or described by more complex S-shaped curves~\cite{binder2008}. For the case of logistic growth, one expects that diffusive particle mixing is not as effective on short time scales as it is in the long time limit.

Contracting media driven by stretched exponential scale factors are also of interest, at least from a methodological point of view. As we have shown in Sect.~\ref{expexp}, an exponential scale factor leads to the onset of a stationary distribution in physical coordinates. It would be interesting to see how this behavior arises as the exponent of the stretched exponential approaches one.
These are just two examples out of the many possibilities that the study of the statistical mechanics of systems with expanding geometries opens.

\section{Acknowledgements}

This work was partially funded by MINECO (Spain) through Grants No. FIS2013-42840-P (partially financed by FEDER funds) (S.~B.~Y. and E.~A.) and MTM2013-40846-P (C.~E.), and by the Junta de Extremadura through Grant No. GR15104 (S.~B.~Y. and E.~A.). We thank Katja Lindenberg for her support and encouragement in the early stages of this work.

\end{document}